\documentclass[aps,ams,amsmath,prx,superscriptaddress,twocolumn]{revtex4-1}
\usepackage{graphics}
\usepackage{graphicx}
\usepackage{epsfig}
\usepackage{amsmath}
\usepackage{amssymb}
\usepackage{amsfonts}
\usepackage{bm}
\usepackage{color}
\usepackage{bbm}
\usepackage{hyperref}
 \usepackage[normalem]{ulem}
\usepackage{comment}
\usepackage{soul}

\renewcommand{\r}{\mathbf{r}}
\newcommand{\q}{\mathbf{q}}

\begin{document}

\title{Giant magnetoresistance in weakly disordered non-Galilean invariant conductors}

\author{Alex Levchenko}
\affiliation{Department of Physics, University of Wisconsin-Madison, Madison, Wisconsin 53706, USA}

\author{Songci Li}
\affiliation{Department of Physics, University of Wisconsin-Madison, Madison, Wisconsin 53706, USA}
\affiliation{Center for Joint Quantum Studies, Department of Physics and Tianjin Key Laboratory of Low-Dimensional
Materials Physics and Preparation Technology, Tianjin University, Tianjin 300354, China}

\author{A. V. Andreev}
\affiliation{Department of Physics, University of Washington, Seattle, Washington 98195, USA}

\date{January 23, 2024}

\begin{abstract}
We develop a hydrodynamic description of electron magnetotransport in conductors without Galilean invariance in the presence of a weak long-range disorder potential. We show that magnetoresistance becomes strong (of order 100~\%) at relatively small fields, at which the inverse square of the magnetic length becomes comparable to disorder-induced variations of the electron density. The mechanism responsible for this anomalously strong magnetoresistance can be traced to the appearance of magnetic friction force in liquids with nonvanishing intrinsic conductivity. We derive general results for the galvanomagnetic and thermomagnetic kinetic coefficients and obtain their dependence on the intrinsic dissipative properties of the electron liquid, and the correlation function of the disorder potential. We apply this theory to graphene close to charge neutrality and cover the crossover to a high-density regime. 
\end{abstract}

\maketitle


\section{Introduction}

The study of electron transport in solids has been a central theme in condensed matter physics. Peierls \cite{Peierls} showed long ago that metallic conductivity can be understood semiclassically, despite the quantum degeneracy of the electron liquid in metals. The reason is that the mean free path of the electrons exceeds their quantum de Broglie wavelength. The semiclassical theory of electron transport \cite{Abrikosov,Gantmakher-Levinson,Pippard} explains the vast majority of transport phenomena in metals and semiconductors. In the simplest version of this theory, the evolution of the electron distribution function is described by the Boltzmann transport equation, in which collisions of electrons with impurities, phonons, and other electrons are described by independent collision integrals. The resulting resistivity is proportional to the sum of partial momentum relaxation rates due to electron-impurity (e-i), electron-phonon (e-ph), and umklapp electron-electron (e-e) collisions. This is in agreement with the phenomenological Matthiessen's rule, which is observed in most metals. 

A departure from this paradigm has been observed recently in several high-mobility systems with low carrier density \cite{Molenkamp,deJong,Gao1,Bandurin-1,Crossno,Ghahari,Morpurgo,Kumar,Bandurin-2,Gusev,Bakarov,Berdyugin,Hone,Brar,Manfra,Hamilton,Gao2,Ponomarenko} and in a variety of materials with strong electron correlations, which are collectively called strange metals \cite{PJH,Mckenzie,Phillips}. In a certain range of temperatures, the rate of umklapp e-e collisions in these systems is negligible, while the rate of momentum-conserving e-e collisions is appreciable and appears to significantly affect the electrical resistivity. 

It is important to note that the apparent independence of resistivity from the rate of momentum-conserving e-e collisions in the framework of the Boltzmann transport equation is not a consequence of the semiclassical approximation, but results from the simplified treatment of disorder and e-e collisions. The description of e-i, e-ph, and e-e collisions by corresponding scattering integrals assumes that these scattering processes are instantaneous, local, and uncorrelated.  In systems where the characteristic spatial scale $\xi$ of the external potential exceeds the mean free path $l_{\mathrm{ee}}$ due to momentum-conserving e-e collisions, the external potential still relaxes electron momentum but this relaxation is significantly affected by the rate of momentum conserving e-e collisions. In this case, the motion of the electron liquid may be described using the hydrodynamic equations~\cite{AKS,Lucas-PRB16,LLA}; see also reviews \cite{Spivak-RMP,Pal,NGMS,Lucas-Fong,ALJS,Narozhny-Review,Fritz-Scaffidi} and references therein.

The first example of hydrodynamic electron transport -- the Poiseuille flow of electron liquid through a clean channel with a rough boundary -- was theoretically considered by Gurzhi~\cite{Gurzhi-UFN} and was subsequently observed in the experiments of Refs.~\cite{Molenkamp,deJong}. A similar situation arises in other finite geometries, such as mesoscopic graphene flakes~\cite{Kumar,Brar}, point contacts~\cite{Guo,Pershoguba,LKL}, and Hall-bar devices~\cite{Crossno,Ghahari,LAL-Hall-bar}. In addition, Corbino geometry has lately attracted significant interest both in experiment \cite{Fuhrer,Faugeras,Zhao-Kim,Hone-Dean,Dietsche,Hakonen} and theory \cite{Shavit,LLA-Corbino,Gall}. It is important to note that in all of these situations the momentum of the electron liquid relaxes via the viscosity-mediated outflow to the sample boundary.

The bulk momentum in the hydrodynamic regime is commonly described by two approaches: (1) momentum relaxation due to short range disorder phenomenologically described by introducing a friction coefficient into the hydrodynamic equations~\cite{Gurzhi-UFN} and (2) long-range disorder, in comparison to the equilibration length, $\xi\gg l_{\text{ee}}$, which results in spatially inhomogeneous flow~\cite{AKS,Lucas-PRB16,LLA}.
In the intermediate regime, where the disorder correlation radius $\xi$ is comparable to $l_{\mathrm{ee}}$, electron transport can be described by the Boltzmann transport equation, provided the disorder is not accounted for by a collision integral, but as an external force modifying the classical trajectories of the electrons~\cite{LMRM}. Note that in the case of long range disorder, the resistivity depends not only on the disorder strength but also on the intrinsic kinetic coefficients of the electron liquid: viscosity and thermal conductivity~\cite{LLA}. 

Magnetotransport in the regime of electron hydrodynamics attracted significant recent attention \cite{Hartnoll,Muller,Muller-Fritz,Mirlin,Alekseev,Xie-Foster,Polini,Scaffidi,Gromov,LXA,PDL,Titov1,Titov2,DasSarma,XL,Mandal,Schutt,Holder,LLA-MR,Gall-MR}. Hydrodynamic effects become especially strong in systems with long range disorder~\cite{LXA,PDL,DasSarma}. In this case the collective character of momentum relaxation in the hydrodynamic regime  significantly affects the magnitude and physical mechanism of magnetoresistance (MR). It arises primarily from the modification of the hydrodynamic flow of the electron liquid by the Lorentz force. The MR becomes strong at relatively weak fields where the cyclotron radius $R_c$ is still much longer than $l_{\mathrm{ee}}$ and therefore bending of electron trajectories by the magnetic field is still negligible. 

The consideration in Ref.~\cite{LXA}  assumed Galilean invariance of the electron liquid. In this case, the current density is given by the momentum density of the electron liquid multiplied by $e/m$, where $m$ is the electron band mass. In the present article, we develop a theory of hydrodynamic magnetotransport in systems, in which the electron liquid does not possess Galilean invariance. We show that  in this case, in addition to modification of the flow by the Lorentz force a qualitatively different mechanism of magnetoresistance arises. Because of the presence of intrinsic conductivity in non-Galilean invariant liquids, the motion of the electron liquid relative to the magnetic field lines produces additional dissipation. 

The goal of this paper is to develop a hydrodynamic theory of electron magnetotransport in systems with long-range disorder, without assuming Galilean invariance of the electron liquid. We obtain general results for the magnetotransport coefficients and apply them to graphene near charge neutrality. 

The paper is organized as follows. In Sec.~\ref{sec:hydro}, we formulate the magnetohydrodynamic description of the electron liquid without Galilean invariance in the presence of a long-range disorder. In Sec.~\ref{sec:linear_response}, we present the linear-response analysis and derive analytical results for the transport coefficients in terms of intrinsic dissipative properties of the electron liquid and the correlation function of the disorder potential. In Sec.~\ref{sec:graphene}, we apply the results to graphene near charge neutrality. In Sec.~\ref{sec:summary}, we summarize the main results. Throughout the paper, we use the natural system of units in which Boltzmann and Planck constants are set to unity $k_B=\hbar=1$.


\section{Magnetohydrodynamic description}\label{sec:hydro}

We consider a two-dimensional (2D) electron system. The hydrodynamic equations describe evolution of the electron liquid at time scales longer than the inelastic relaxation time $\tau_{\text{ee}}$ due to electron-electron (e-e) collisions and spatial scales longer than the corresponding relaxation length $l_{\text{ee}}$. At such scales the liquid comes to a local thermal equilibrium and its state is described by the densities of quantities, which are conserved by e-e collisions, i.e., the densities of particles $n$, energy $\epsilon$, and momentum ${\bf p}$. Therefore, hydrodynamic equations describe evolution of $n$, $\epsilon$, and ${\bf p}$. Due to local equilibrium of the electron liquid,  the evolution equation for energy density can be replaced by the equivalent evolution equation for the entropy density $s$~\cite{LL-V6}. In this form the hydrodynamic equations for an electron liquid moving in a static external potential $U(\mathbf{r})$ become
\begin{subequations}\label{eq:continuity}
\begin{eqnarray}
&& 
\partial_t n= - \boldsymbol{\nabla}\cdot \bf j,  \label{eq:continuity_n} \\
&& \partial_t {\bf p} = - \bm{\nabla} \cdot \hat{ \Pi} -n\, \bm{\nabla} (e\phi + U) + \frac{e}{c} \, [{\bf j} \times {\bf H}], \label{eq:continuity_p}\\
&& \partial_t s=- \boldsymbol{\nabla}\cdot {\bf j}_s +\varsigma.\label{eq:s_dot_source}
\end{eqnarray}
\end{subequations}
Here and in what follows, we denote vector quantities by bold-face symbols and denote Cartesian indices by Latin subscripts. In Eq. \eqref{eq:continuity_p} $\phi$ is the electric potential related to the electron density by the Poisson equation and $\hat{\Pi}$ denotes the momentum flux  tensor of the electron liquid. The last term of Eq.~(\ref{eq:continuity_p}) describes the Lorentz force exerted on the liquid by the applied magnetic field ${\bf H}$. Below, we focus on the stationary magnetotransport and assume that the magnetic field is applied along the positive $z$-axis, which is normal to the sample plane, ${\bf H} = H \hat{\bf z}$. In Eq. \eqref{eq:s_dot_source}, ${\bf j}_s$ is the entropy flux and $\varsigma$ is the local rate of entropy production due to e-e collisions.

The fluxes of particles ${\bf j}$, entropy ${\bf j}_s$, and momentum $\hat{\Pi}$ may be separated into the equilibrium (convective) and nonequilibrium (dissipative) components, 
\begin{subequations}\label{eq:constitutive}
\begin{eqnarray}
    &&{\bf j} = n {\bf u} + {\bf j}', \label{eq:constitutive_j}\\
    &&{\bf j}_s = s {\bf u} + {\bf j}'_s, \label{eq:constitutive_j_epsilon} \\
    &&\Pi_{ij} = P \delta_{ij}  -\sigma'_{ij}, \label{eq:constitutive_Pi}
\end{eqnarray}
\end{subequations}
where ${\bf u}$ and $P$ are, respectively, the local hydrodynamic velocity and pressure. The first terms in the right hand side in the above equations denote the equilibrium components of fluxes of conserved quantities. The primed quantities denote the dissipative parts of the fluxes of conserved quantities and are proportional to gradients of the equilibrium parameters: ${\bf u}$, $T$, and $P$. 

The viscous stress tensor $\sigma'_{ij}$ in Eq.~(\ref{eq:constitutive_Pi}) is given by \cite{LL-V6}
\begin{equation}\label{eq:viscous_stress}
\sigma_{ij}' = \eta (\partial_i u_j + \partial_j u_i) + \left(\zeta-\eta\right) \delta_{ij}
\partial_k u_k,
\end{equation}
where $\eta$ and $\zeta$ are, respectively, shear and bulk viscosities. 

To make the presentation more compact, we combine the particle and entropy fluxes into a two-component column vector
\begin{equation}\label{eq:J_column}
  \vec{\bf J}= \left(
                      \begin{array}{l}
                        \bf j \\
                        {\bf j}_s \\
                      \end{array}
                    \right).
\end{equation}
We denote the column-vector quantities by arrows above them and use boldface letters to denote the usual spatial vectors. Following the notations of Ref.~\cite{LL-V5}, we denote densities of thermodynamic variables by $\vec{x}$ and combine the particle and entropy densities into a column vector
\begin{equation}\label{eq:_little_x_def}
 \vec{x}= \left(
    \begin{array}{c}
      n \\
      s \\
    \end{array}
  \right).
\end{equation}
In these notations the constitutive relations Eqs. (\ref{eq:constitutive_j}) and (\ref{eq:constitutive_j_epsilon}) for the particle and entropy currents can be written in the form
\begin{eqnarray}\label{eq:column_constitutive}
 \vec{\bf J}= \vec{x}{\bf u} - \hat{\Upsilon}\vec{\bf X}',
\end{eqnarray}
where $\hat{\Upsilon}$ is the matrix of intrinsic kinetic coefficients, which  characterizes the dissipative properties of the electron liquid, and the column vector $\vec{\bf X}'$ represents the thermodynamic forces conjugate to the variables $\vec{x}$~\cite{LL-V5}. The column vector of thermodynamic forces consists of the electromotive force (EMF) and the temperature gradient, and can be written in the form \cite{LL-V8}
\begin{equation}\label{eq:X'_def}
  \vec{\bf X}'= \vec{\bf X}   - \frac{e}{c} [\bf u \times \bf H] \left(
    \begin{array}{c}
      1 \\
       0 \\
    \end{array}
  \right).
\end{equation}
The second term above describes the contribution to EMF caused by the Lorentz force exerted by a magnetic field in a moving liquid \cite{Note}. 
The first term can be written as a pure gradient and is given by 
\begin{equation}\label{eq:X_def}
\vec{\bf X}= \left(
    \begin{array}{c}
      - e\boldsymbol{\mathcal{E}} \\
       \boldsymbol{\nabla} T \\
    \end{array}
  \right).
\end{equation}
Here $e  \boldsymbol{\mathcal{E}}$ represents the potential part of the EMF given by the gradient of electrochemical potential 
\begin{equation}\label{eq:EMF_def_n}
  e  \boldsymbol{\mathcal{E}}\equiv -\boldsymbol{\nabla} \left( \mu + e\phi + U\right),
\end{equation}
with $\mu$ being the chemical potential. The appearance of the hydrodynamic velocity in the electromotive force in Eq.~\eqref{eq:X'_def}  corresponds to evaluating the electric field in the reference frame moving with the liquid.  

The  entropy production rate $\varsigma$ in Eq.~(\ref{eq:s_dot_source}) can be expressed in terms of  the viscous stresses and  thermodynamic force $\vec{\bf X}'$ in the form
\begin{equation}\label{eq:s_dot_X}
\varsigma=\frac{1}{2}\int \sigma'_{ij} (\partial_i u_j+ \partial_j u_i) d\mathbf{r}+\int \vec{\bf X}'^{\mathbb{T}} \hat{\Upsilon}\vec{\bf X}'d\mathbf{r}. \end{equation}
The superscript $\mathbb{T}$ denotes transposition and integration extends over the two-dimensional system area. 

We discuss next the matrix $\hat{\Upsilon}$ of kinetic coefficients of the electron liquid introduced in Eq.~\eqref{eq:column_constitutive}. We consider the regime of semiclassically weak magnetic fields, where $l_{\text{ee}}$ is shorter than the cyclotron radius $R_c$ of a typical thermal electron. For example, for electrons in single layer graphene near charge neutrality, this is justified provided the magnetic length $l_H = \sqrt{c/e H}$ exceeds the thermal de Broglie wavelength $l_T=v/T$. In this regime the kinetic coefficient of the liquid may be assumed to be equal to their values at $H=0$. With this accuracy the liquid has vanishing intrinsic Hall response. Thus the intrinsic conductivity $\sigma$, the intrinsic thermal conductivity $\kappa$, and the intrinsic thermoelectric coefficient $\gamma$ have purely scalar character. The $2\times 2$ matrix  $\hat{\Upsilon}$ of kinetic coefficients in the column-vector space is given as follows \cite{Aleiner}:
\begin{equation}
\label{eq:gamma_def}
\quad \hat{\Upsilon}= 
\left(\begin{array}{cc}
\sigma/e^2 & \gamma/T \\
\gamma/T & \kappa/T  \\
\end{array}
\right).
\end{equation}

In linear response, we may neglect the entropy production and set $\varsigma \to 0$ in Eq.~(\ref{eq:s_dot_source}). Then in the stationary case the continuity equation (\ref{eq:continuity_n}) and the entropy evolution equation (\ref{eq:s_dot_source}) reduce to the continuity equations for the column vector current. Using Eqs.~(\ref{eq:column_constitutive}) and (\ref{eq:X'_def}) the latter can be expressed in the form
\begin{equation}\label{eq:column_current_continuity_n}
 \boldsymbol{\nabla} \cdot\left( \left[ \vec{x}   + \frac{1}{l_H^2}  \vec{\gamma}  \hat{\mathbf{z}} \times \right]{\bf u} - \hat{\Upsilon}{\vec{\bf X}} \right)=0.
\end{equation}
Here we introduced a shorthand notation
 \begin{equation}\label{eq:vect_gamma_n}
  \vec{\gamma} = \left(
                   \begin{array}{c}
                     \frac{\sigma}{e^2} \\
                     \frac{\gamma}{T} \\
                   \end{array}
                 \right).
\end{equation}
Using Eqs.~(\ref{eq:constitutive_Pi}), (\ref{eq:viscous_stress}), and the thermodynamic relation
\begin{equation}
    \label{eq:thermo_relation}
    \boldsymbol{\nabla} P = n \boldsymbol{\nabla} \mu + s \boldsymbol{\nabla} T,
\end{equation}
the momentum balance equation (\ref{eq:continuity_p}) may be expressed in the column vector notations as
\begin{equation}\label{eq:force_balance_column_H_n}
\left[ \vec{x}  + \frac{1}{l_H^2} \vec{\gamma}\, \hat{\bf z} \times \right]^{\mathbb{T}}\vec{\bf X}' = 
   \left( \bm{\nabla} \cdot   \eta \bm{\nabla} \right)\mathbf{u} + \bm{\nabla} \left[\zeta \left(\boldsymbol{\nabla} \cdot \mathbf{u}\right)\right]
\end{equation}

The system of hydrodynamic equations (\ref{eq:continuity}) and constitutive relations (\ref{eq:constitutive}), together with Eq.~\eqref{eq:column_constitutive}, presented in this section does not assume Galilean invariance and provides a general description of the flow of electron liquid in an external random potential at small velocities. For the Galilean-invariant liquids $\vec{\bf J}= \vec{x}{\bf u}$ and $\sigma=\gamma=0$. In this case Eq. \eqref{eq:s_dot_X} reproduces the well known result for the energy dissipation rate in Galilean-invariant liquids \cite{LL-V6} and a textbook form of Eq. \eqref{eq:force_balance_column_H_n} for the force balance condition \cite{LL-V8}. 

Within the accuracy of our approximation, the effect of the magnetic field on thermoelectric transport arises from two different mechanisms: (i) modification of the flow by the Lorentz force in Eq.~\eqref{eq:continuity_p} (this changes the convective parts of charge and heat currents) and (ii) the fact that the EMF appearing in Eqs.~(\ref{eq:column_constitutive}) and (\ref{eq:X'_def}) also contains the Lorentz force (this changes the dissipative parts of charge and heat currents). For Galilean-invariant liquids, which are characterized by the vanishing intrinsic conductivity $\sigma$ and thermoelectric coefficient $\gamma$, only the first of these mechanisms is at play. Hydrodynamic magnetoresistance for Galilean-invariant electron liquids was studied in Ref.~\cite{LXA}. We show below that the second mechanism arising in the absence of Galilean invariance leads to new effects, which become especially pronounced at charge neutrality. Therefore we expect that the predictions of the theory developed below may be tested in graphene devices at low carrier density. 

We close this section by noting that the approximate form of the magnetohydrodynamic (MHD) equations for weakly conducting liquids in the absence of disorder was originally derived by Braginskii \cite{Braginskii}. The theory presented in the present work is analogous to that in the sense that the effect of the electric currents on the magnetic field can be neglected. In this approximation the magnetic field lines cannot be entrained by the liquid. This represents a key difference with magnetohydrodynamics of ideally conducting fluids, in which the magnetic field lines are frozen into the liquid by Alfv\'en's theorem \cite{Alfven}. Therefore, the Maxwell equations describing the evolution of the magnetic field are not needed, which simplified the treatment considerably.  In this regime the motion of the liquid relative to the magnetic field lines produces a magnetic friction force, which is proportional to the intrinsic conductivity of the liquid. As we show below, this force has a significant effect on magnetoresistance near charge neutrality.


\section{Electron magnetotransport at weak disorder}\label{sec:linear_response}

We now use the description presented in Sec.~\ref{sec:hydro} to study electron transport in the linear response regime in a disordered system. 
The technical approach is the same as reported in our earlier work \cite{LLA}. Here we present generalizations needed to account for the effect of finite magnetic field. 

First, recall that, in thermal equilibrium, the hydrodynamic velocity vanishes. Since both the temperature and the electrochemical potential $\mu + e\phi + U$ are spatially uniform the thermodynamic force $\vec{\bf X}$ vanishes. As a result the force balance condition is trivially satisfied. Away from equilibrium one needs to find a nonvanishing spatial distribution of the hydrodynamic velocity $\mathbf{u} (\bf{r})$ and the thermodynamic force $\vec{\bf X}(\bf{r})$ that solve the system of equations (\ref{eq:column_current_continuity_n}) and (\ref{eq:force_balance_column_H_n}). In linear response the number density and entropy density in the column vector $\vec{x}(\bf{r})$, as well as the dissipative coefficients of the liquid, $\hat{\Upsilon}(\bf{r})$, $\eta(\bf{r})$, and $\zeta(\bf{r})$, are given by their equilibrium values. Once the solutions ${\bf u}(\bf r)$ and $\vec{\bf X}(\bf r)$ are found, we obtain the macroscopic densities and entropy flux via the relation
\begin{equation}\label{eq:averaged_current_spatial}
  \langle \vec{\mathbf{J}}\rangle=\left\langle \vec{x} (\mathbf{r}) \mathbf{u} (\mathbf{r}) - \hat{\Upsilon} (\mathbf{r})  \vec{\mathbf{X}}' (\mathbf{r}) \right\rangle,
\end{equation}
where $\langle \ldots \rangle \equiv \frac{1}{A} \int d \mathbf{r} \ldots $ denotes spatial average over the system area $A$.

In the present approach the disorder potential manifests itself via the spatial dependence of the equilibrium number and entropy densities, $n(\bf{r})$ and $s(\bf{r})$,  the matrix of kinetic coefficients,  $\hat{\Upsilon}(\bf{r})$, and the viscosities $\eta(\bf{r})$ and $\zeta(\bf{r})$. To further simplify our discussion, we assume that these quantities are weakly inhomogeneous and use perturbation theory in disorder to derive analytical results for the macroscopic thermoelectric conductivity matrices $\hat{\Upsilon}$ and $\hat{\Xi}$. The latter link the macroscopic fluxes $\langle \vec{\bf J} \rangle = \langle ({\bf j},{\bf j}_s)^{\mathbb{T}}\rangle$ to the average electric field and the temperature gradient in the system $\vec{\bf X}_0\equiv \langle \vec{\bf X} \rangle$:
\begin{equation}\label{eq:Upsilon_eff_def}
   \langle \vec{\mathbf{J}}\rangle  = -\hat{\Upsilon} \vec{\mathbf{X}}_0-\hat{\Xi} (\hat{\mathbf{z}}\times \vec{\bf X}_0).
\end{equation}
As explained in Sec. \ref{sec:hydro}, we neglect the dependence of the kinetic coefficients of the liquid on the magnetic field as the correction to the transport coefficient is small in the hydrodynamic regime in a parameter of a ratio $l_{\mathrm{ee}}/\xi\ll1$.

\subsection{Perturbation theory in disorder}
\label{sec:perturbation}

We express the densities of conserved quantities $\vec{x}(\r)$ and intrinsic kinetic coefficients $\hat{\Upsilon}(\r)$ describing the equilibrium state of the liquid in the  form
\begin{equation}
\vec{x}(\r) = \vec{x}_0 +\delta \vec{x}(\r), \quad  \hat{\Upsilon}(\r) = \hat{\Upsilon}_0 + \delta \hat{\Upsilon}(\r),
\end{equation}
where the uniform components are denoted by the subscript $0$ and the spatial variations are small: $\delta \vec{x}(\r) \ll \vec{x}_0$ and $\delta \hat{\Upsilon}(\r) \ll \hat{\Upsilon}_0$. The resulting hydrodynamic velocity $\mathbf{u}(\r)$ and the thermodynamic force $\vec{\bf X} (\r)$ are thus also weakly inhomogeneous,
\begin{equation}
{\bf u}(\r) = {\bf u}_0 +\delta {\bf u}(\r), \quad  \vec{\bf X} (\r) = \vec{\bf X}_0 + \delta \vec{\bf X} (\r),
\end{equation}
with $\delta {\bf u}(\r) \ll  {\bf u}_0$ and $\delta \vec{\bf X} (\r) \ll \vec{\bf X}_0$.

We seek the solutions to $\delta {\bf u}(\r)$ and the driving force $\delta \vec{\bf X} (\r)$ as perturbations to their uniform counterparts. The solution scheme can be summarized as follows. Working in the Fourier representation, we express the inhomogeneous part of the flow velocity field $\delta {\bf u}$ and forces $\delta \vec{\bf X}$ in terms of their uniform counterparts from the continuity equation for the column current, Eq.~(\ref{eq:column_current_continuity_n}), and the momentum balance equation, Eq.~(\ref{eq:force_balance_column_H_n}). This enables us to express the spatial average of the currents in terms of $\vec{\bf{X}}_0$ and ${\bf u}_0$ with the aid of Eq.~(\ref{eq:averaged_current_spatial}). Furthermore, the spatial average of the momentum balance equation (\ref{eq:force_balance_column_H_n}) imposes a linear relation between $\vec{\bf{X}}_0$ and ${\bf u}_0$. Finally, with this relation we express the macroscopic current $\vec{\bf J}$ in the form of Eq.~(\ref{eq:Upsilon_eff_def}) and extract transport  coefficients. 

To implement this scheme, it is convenient to work in the Fourier representation by defining the Fourier amplitude of a given quantity $O (\mathbf{r})$ in the standard way,
\begin{equation}\label{eq:Fourier_def}
  O_{\mathbf{q}} = \int d\r\, O (\r)\,  \mathrm{e}^{- i \mathbf{q} \cdot \r}.
\end{equation}
To the first order in perturbation in the inhomogeneity, Eq.~(\ref{eq:force_balance_column_H_n}) yields the following condition for the Fourier harmonics with nonzero wave vector:
\begin{align}
&0 =\eta q^2 {\bf u}_\q
+ \zeta  \q (\q \cdot {\bf u}_\q)
\nonumber \\ 
&+ \left[ \frac{\sigma_0}{l_H^4 e^2 } -  \frac{n_0}{l_H^2} \hat{\bf z} \times  \right]  {\bf u}_ \q +
\left[ \vec{x}_0^{\mathbb{T}}  + \frac{1}{l_H^2}   \vec{\gamma}_0^{\mathbb{T}} \hat{\bf z} \times \right]\vec{\bf X}_\q 
\nonumber \\ 
&+\left[ \vec{x}_\q^{\mathbb{T}}  
+ \frac{1}{l_H^2}   \vec{\gamma}_\q^{\mathbb{T}} \hat{\bf z} \times \right]  \vec{\bf X}_0 + \left[ \frac{\sigma_\q}{l_H^4 e^2 } -  \frac{n_\q}{l_H^2} \hat{\bf z} \times  \right]  {\bf u}_ 0. \label{eq:u_q_X_column_H_n}
\end{align}
The continuity equation (\ref{eq:column_current_continuity_n}) for particle and entropy currents is given by
\begin{align}\label{eq:continuity_perturbation_H_n}
  0 =  \vec{x}_0 (\q \cdot {\bf u}_\q)-\hat{\Upsilon}_0\, (\q \cdot \vec{\bf X}_\q)
    - \frac{1}{l_H^2}
    \vec{\gamma}_0 \,   (\hat{\bf z} \cdot [ \q \times {\bf u}_\q ])\nonumber \\
    + \,
   \vec{x}_\q (\q \cdot {\bf u}_0)    - \frac{1}{l_H^2} \vec{\gamma}_\q \, (\hat{\bf z} \cdot  [ \q \times {\bf u}_0]) -
   \hat{\Upsilon}_\q\, (\q \cdot \vec{\bf X}_0).
\end{align}
The system of Eqs.~(\ref{eq:u_q_X_column_H_n}) and (\ref{eq:continuity_perturbation_H_n}) determines the inhomogeneous components of the hydrodynamic velocity, ${\bf u}_\q$,  and the thermodynamic forces, $\vec{\bf X}_\q$, in terms of their macroscopic counterparts, ${\bf u}_0$ and $\vec{\bf X}_0$. However, since the macroscopic flow is characterized by only two macroscopic currents (particle and entropy flux, $\langle {\bf j} \rangle, \langle {\bf j}_s \rangle$), the average velocity ${\bf u}_0$ is determined by the macroscopic  thermodynamic forces $\vec{\bf X}_0$.  The relation between them can be obtained by considering the uniform Fourier component of the  force balance equation (\ref{eq:force_balance_column_H_n}), which can be cast in the form
\begin{equation}\label{eq:force_balance_H_average_n}
  \langle \vec{x}^{\mathbb{T}} \vec{\bf X} \rangle +  \frac{1}{l_H^2} \hat{\bf z} \times \langle \vec{\gamma}^{\mathbb{T}} \vec{\bf X} \rangle + \frac{1}{l_H^4} \,  \frac{\langle \sigma  \bf u \rangle}{e^2}  -  \frac{1}{l_H^2} \hat{\mathbf{z}} \times \langle n \mathbf{u} \rangle = 0.
\end{equation}

In order to find the solutions for $\bf u_{\bf q}$ and $\vec{\bf X}_{\bf q}$ from Eqs.~(\ref{eq:u_q_X_column_H_n}) and (\ref{eq:continuity_perturbation_H_n}), it is useful to decompose them into longitudinal ($l$) and transverse components ($t$) relative to $\bf q$. Notice that in two dimensions we can decompose all vector quantities into the longitudinal and transverse parts as follows,
\begin{equation}\label{eq:t_l_decomposition_n}
  {\bf V}_\q  = {\bf V}^l_\q + {\bf V}^t_\q = \frac{\q}{q} V^l_\q + \frac{\hat{\bf{z}} \times \q}{q} V^t_\q,
\end{equation}
where
\begin{eqnarray}\label{eq:long_transverse_def_n}
 && V^l_\q = \frac{1}{q}  \, (\q \cdot {\bf V}_\q ), \qquad V^t_\q = \frac{1 }{q} \,  (\hat{\bf{z}} \cdot [\q \times  {\bf V}_\q]).
\end{eqnarray}
Taking this into account, the solution of the system of linear Eqs.~(\ref{eq:u_q_X_column_H_n}) and (\ref{eq:continuity_perturbation_H_n})  is given by
\begin{subequations}\label{eq:solution_summary}
\begin{align}
 \vec{X}^l_\q = & \, \frac{\hat{\Upsilon}_0^{-1} \vec{x}_0^{\mathbb{T}} \otimes  \vec{x}_0  \hat{\Upsilon}_0^{-1}  - \lambda_q^H  \hat{\Upsilon}_0^{-1} }{\lambda_q^H}
     \hat{\Upsilon}_\q  \vec{X}^l_0 \nonumber \\
& + \frac{1}{ l_H^2  } \hat{\Upsilon}_0^{-1}  \left(  \frac{1}{\eta q^2   }
     \vec{\gamma}_0  \otimes   \vec{x}_\q^{\mathbb{T}}      + \frac{1}{\lambda_q^H  }  \vec{x}_0 \otimes \vec{\gamma}_\q^{\mathbb{T}} \right) \vec{X}^t_0 \nonumber \\
&+ \frac{ \left( \lambda_q^H  - \frac{\sigma_\q}{e^2 l_H^4}  \right)  \hat{\Upsilon}_0^{-1} - \hat{\Upsilon}_0^{-1}  \vec{x}_0 \otimes  \vec{x}_0^{\mathbb{T}}  \hat{\Upsilon}_0^{-1}    }{\lambda_q^H}\vec{x}_\q u_0^l \nonumber \\ 
&- \frac{1}{\lambda_q^H}    \hat{\Upsilon}_0^{-1} \vec{x}_0^{\mathbb{T}} \otimes   \vec{x}_\q^T \vec{X}^l_0 
- \frac{1}{l_H^2}
     \hat{\Upsilon}_0^{-1} \vec{\gamma}_\q u^t_0. \label{eq:X_q_result} \\  
 u_\q^t = & \, 
  - \frac{1}{ \eta q^2}  \, \vec{x}_\q^{\mathbb{T}}\,   \vec{X}^t_0, \label{eq:u_transverse_summary_sol} \\
 u_\q^l= & \, - \left(\vec{x}_0^{\mathbb{T}} \,
 \hat{\Upsilon}_0^{-1}      \vec{x}_\q+\frac{\sigma_\q}{e^2 l_H^4}\right)
 \, \frac{u^l_0}{\lambda^H_q}\nonumber \\ 
 &+ \left(
\vec{x}_0^{\mathbb{T}}  \hat{\Upsilon}_0^{-1}  \hat{\Upsilon}_\q
- \vec{x}_\q^{\mathbb{T}}
\right)
\, \frac{\vec{X}^l_0}{\lambda^H_q}+\frac{\vec{\gamma}^{\mathbb{T}}_\q  \vec{X}^t_0}{\lambda^H_q l_H^2} .\label{eq:u_longitudinal_summary_sol}
\end{align}
\end{subequations}
Here $\lambda^H_q$ is given by
\begin{equation}\label{eq:lambda_q_H_def} 
\lambda^H_q = \lambda_q+\frac{\sigma_0}{e^2 l_H^4}, \qquad \lambda_q=(\eta + \zeta) q^2 +\vec{x}_0^{\mathbb{T}} \,
 \hat{\Upsilon}_0^{-1} \vec{x}_0.
\end{equation} 
In the above expressions we used the standard notation for the direct product of two vectors $\vec{a}\otimes\vec{b}^{\mathbb{T}}$ that defines a corresponding matrix. From the obtained solution one should note that the transverse (vortical) component of the velocity is unaffected by the magnetic field. This assumes that we take $\vec{X}$ and the wave vector $\mathbf{q}$ to be independent of the magnetic field. 

Next we simplify the general expressions obtained above for the experimentally relevant case of long range disorder. Specifically, when the correlation length of disorder satisfies 
\begin{equation}\label{eq:large_correlation_length}
\varepsilon\equiv\frac{1}{\xi^2}\frac{\eta+\zeta}{\vec{x}_0^{\mathbb{T}} \,
 \hat{\Upsilon}_0^{-1} \vec{x}_0} \ll 1,
\end{equation}
we can drop the $q$ dependence in $\lambda^H_q$. In accordance with the long-range disorder condition, the number density satisfies the following constraint:
\begin{equation}\label{eq:n_constraint}
	n^2_0 \lesssim \frac{k\xi^2\lambda_0}{\sqrt{\eta(\eta+\zeta)}}.
\end{equation}
It is worthwhile to notice that, under the condition of Eq.~(\ref{eq:large_correlation_length}), the hydrodynamic flow is primarily of a vortical character (the magnitude of $u^t$ exceeds that of $u^l$) on  spatial scales of order $\xi$. Furthermore, under a weak magnetic field, $l^2_Hs_0\gg 1$, the field dependence in $\lambda^H_q$ is also insignificant. Then we get simplified expressions
\begin{subequations}\label{eq:solution_summary_1}
\begin{align}
 \vec{X}^l_\q = & \, {\hat K}\left(\vec{x}_\q u_0^l-\hat{\Upsilon}_\q  \vec{X}^l_0\right)      - \frac{1}{\lambda_0}    \hat{\Upsilon}_0^{-1} \vec{x}_0^{\mathbb{T}} \otimes   \vec{x}_\q^T \vec{X}^l_0\nonumber \\ 
&+ \frac{1}{ l_H^2  } \hat{\Upsilon}_0^{-1}    \frac{1}{\eta q^2   }
     \vec{\gamma}_0  \otimes   \vec{x}_\q^{\mathbb{T}}       \vec{X}^t_0 - \frac{1}{l_H^2}
     \hat{\Upsilon}_0^{-1} \vec{\gamma}_\q u^t_0, \label{eq:X_q_result_1} \\
 u_\q^t = & \, 
  - \frac{1}{ \eta q^2}  \, \vec{x}_\q^{\mathbb{T}}\,   \vec{X}^t_0,  \label{eq:u_transverse_summary_sol_1} \\
 u_\q^l= & \, - \left(\vec{x}_0^{\mathbb{T}} \,
 \hat{\Upsilon}_0^{-1}      \vec{x}_\q+\frac{\sigma_\q}{e^2 l_H^4}\right)
 \, \frac{u^l_0}{\lambda_0}\nonumber \\ 
&+ \left(
\vec{x}_0^{\mathbb{T}}  \hat{\Upsilon}_0^{-1}  \hat{\Upsilon}_\q
- \vec{x}_\q^{\mathbb{T}}
\right)
\, \frac{\vec{X}^l_0}{\lambda_0}+\frac{\vec{\gamma}^{\mathbb{T}}_\q  \vec{X}^t_0}{\lambda_0 l_H^2} , \label{eq:u_longitudinal_summary_sol_1}
\end{align}
\end{subequations}
where we introduced the matrix
\begin{equation}
\label{eq:K_hat}
 \hat{K}  = \hat{\Upsilon}_0^{-1} - \frac{1}{\lambda_0} \hat{\Upsilon}_0^{-1}   \vec{x}_0\otimes  \vec{x}_0^{\mathbb{T}} \hat{\Upsilon}_0^{-1}.
\end{equation}
It is clear that $u^l_{\q} \ll u^t_{\q}$ for the long wavelength fluctuations, $q\to0$. In this approximation, which can be termed as the incompressible liquid approximation, substituting Eqs.~\eqref{eq:X_q_result_1} and~\eqref{eq:u_transverse_summary_sol_1} into the force balance equation (\ref{eq:force_balance_H_average_n}) yields the following relation between $\mathbf{u}_0$ and $\vec{\bf X}_0$:
\begin{align}\label{eq:u_0_X_0_H}
 &\vec{x}_e^{\mathbb{T}} \vec{\bf X}_0 +  \frac{1}{l_H^2}  [\hat{\bf{z}} \times  \vec{\gamma}_0^{\mathbb{T}} \vec{\bf X}_0]= \nonumber \\ 
 &- \left( k +  \frac{1}{l_H^4} \,  \frac{ \sigma_0  }{e^2} \right) {\bf u}_0 +  \frac{n_0 }{l_H^2}  [\hat{\bf{z}} \times {\bf u}_0]. 
\end{align}
 Here the effective friction coefficient can be expressed in terms of the matrix $\hat{K}$
from Eq. (\ref{eq:K_hat})
 \begin{equation}\label{eq:friction_coefficient}
  k = \frac{1}{2}\int_{\bf{q}}  \vec{x}_{-\q}^{\mathbb{T}}  \hat{K} \vec{x}_\q. 
\end{equation}
The disorder-renormalized density $\vec{x}_{\rm{e}}$ in the first term on the left hand side of Eq.~(\ref{eq:u_0_X_0_H}) is given by
\begin{equation}\label{eq:x_e_long_wavelength}
  \vec{x}_{\rm{e}} = \vec{x}_0 - \frac{1}{2} \left\langle \delta \hat{\Upsilon} \hat{K} \delta \vec{x}  \right\rangle    - \frac{1}{2 \lambda_{0}} \left\langle  \delta \vec{x} \otimes \delta \vec{x}^{\mathbb{T}} \right\rangle   \hat{\Upsilon}^{-1}_0 \vec{x}_0.
\end{equation}
Equation \eqref{eq:u_0_X_0_H} expresses the force balance condition. The first term on the left hand side (LHS) corresponds to the force exerted on the liquid by the external electric field and thermally induced pressure gradient. The second term on the right hand side (RHS) is  the Lorentz force due to the flow. Finally, the second term on the LHS and the first term on the RHS describe the two friction  forces that act on the liquid.  The first one is the disorder-induced friction force $-k {\bf u}_0$ whose magnitude is proportional to the disorder strength and the velocity in the laboratory frame. The second one is the magnetic friction force. Substituting Eq.~\eqref{eq:vect_gamma_n} into the second term on the RHS it is easy to see that  this force has the form $-k_H\left( \mathbf{u}_0 - c \, \frac{\mathbf{E}\times \mathbf{H}}{H^2}\right)$.   It tries to bring the liquid to the frame moving with velocity $ c \, \frac{\mathbf{E}\times \mathbf{H}}{H^2}$ of the magnetic field lines. The corresponding friction coefficient,
\begin{equation}
k_H=\frac{ \sigma_0}{e^2l_H^4}, 
\end{equation}
is quadratic in the magnetic field and is proportional to the intrinsic conductivity of the electron liquid. In the limiting case of an ideally conducting liquid the friction force diverges and one recovers the result of Alfv\'en's theorem on frozen-in magnetic field lines in magnetohydrodynamics \cite{LL-V8,Alfven}, according to which the motion of a perfectly conducting liquid perpendicular to the magnetic field lines is forbidden.  

In a steady state these two friction forces are balanced by the external force acting on the liquid. 
From Eq.~(\ref{eq:u_0_X_0_H}) we find as a result 
\begin{equation}\label{eq:u_0_X_0_H_result}
{\bf u}_0 = - \frac{  \left[
(k+k_H)  +  \frac{n_0 }{l_H^2}  \hat{\bf{z}} \times
\right] \left[  \vec{x}_e^{\mathbb{T}} \vec{\bf X}_0 +  \frac{1}{l_H^2}  \hat{\bf{z}} \times  \vec{\gamma}_0^{\mathbb{T}} \vec{\bf X}_0 \right]}{(k+  k_H)^2 +n_0^2/l_H^4}.
\end{equation}
Note that in this expression the denominator changes significantly while the magnetic field is still small, $l_H^{-4} \sim  \langle \delta n^2 \rangle^2/ \left[  \langle \delta n^2 \rangle + n_0^2\right] $,  and still satisfies our assumptions. Also it can be readily verified that, in the clean limit, namely taking $k\to 0$ and simultaneously $\vec{x}_{\mathrm{e}}\to\vec{x}_0$, one reproduces the corresponding hydrodynamic velocity in the pristine system.  

\begin{figure*}[t!]
  \centering
  \includegraphics[width=0.45\linewidth]{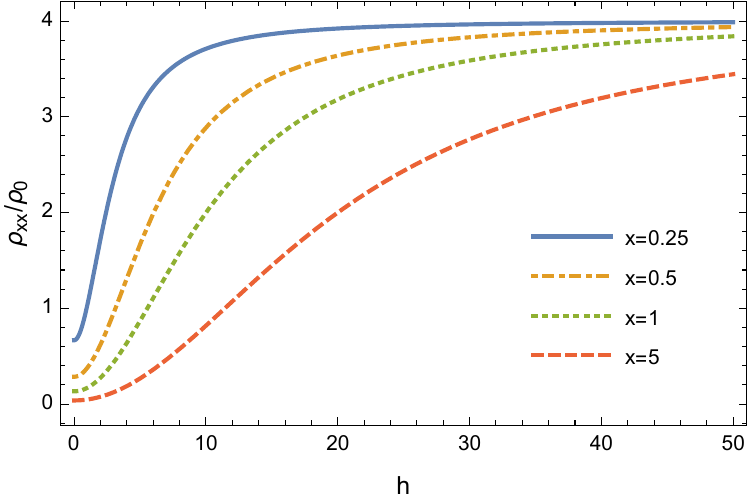}
  \includegraphics[width=0.45\linewidth]{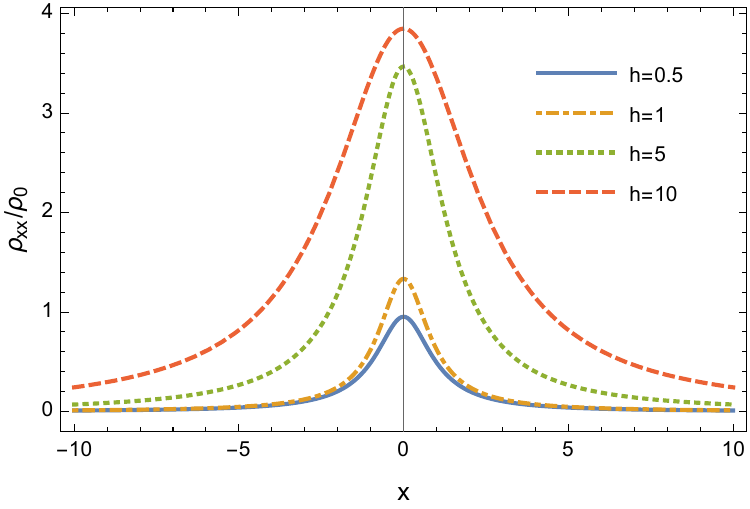}
   \includegraphics[width=0.45\linewidth]{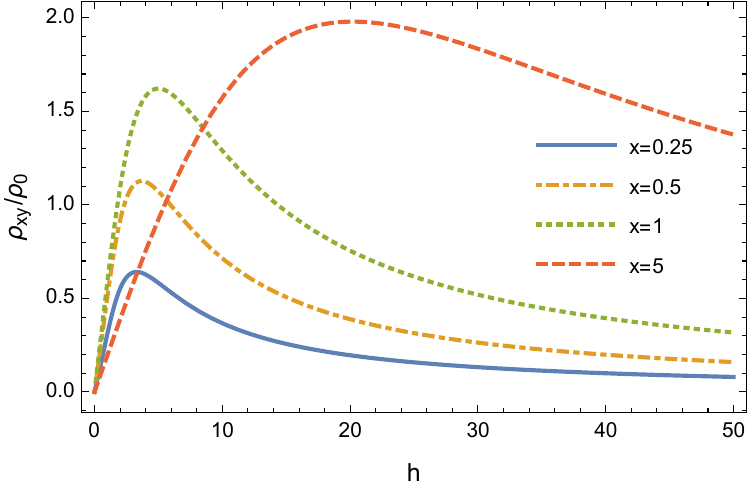}
  \includegraphics[width=0.45\linewidth]{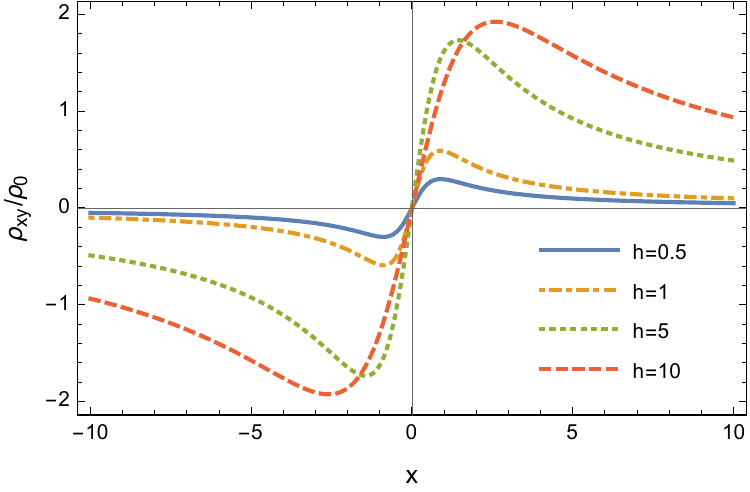}
  \caption{Magnetic field and density dependence for the magnetoresistance $\rho_{xx}(n_0,H)$ and the Hall resistances $\rho_{xy}(n_0,H)$ plotted for a particular choice of a single dimensionless parameter  $\chi_\sigma=e^2\chi/\sigma_0=0.25$. Both magnetoresistivities were normalized in units of intrinsic resistivity $\rho_0=\sigma^{-1}_0$. Magnetic field was normalized by a natural scale in the problem, namely $h= \frac{\sigma_0}{e^2 l_H^2}\sqrt{2/\langle \delta n^2\rangle}\propto H$, while density is plotted in units of $x=\sqrt{2}n_0/\sqrt{\langle\delta n^2\rangle}$ [Eq. \eqref{eq:chi}].}
  \label{Fig-r-xx-xy}
\end{figure*}

\subsection{Macroscopic conductivity matrix}

The macroscopic current, given by Eq. (\ref{eq:averaged_current_spatial}), can be written in the form
\begin{align}
\langle \vec{\bf J} \rangle  = & \, \left\langle \vec{x}{\bf u} - \hat{\Upsilon}\vec{\bf X}   -   \vec{\gamma} \,  \frac{e}{c} [\bf H \times \bf u] \right\rangle  \nonumber \\
= &\, \left[  \vec{x}_0  + \frac{1}{l_H^2}   \vec{\gamma}_0 \,  \hat{\bf z} \times  \right] {\bf u}_0     - \hat{\Upsilon}_0\vec{\bf X}_0 
   +\langle \delta  \vec{x} \delta {\bf u}\rangle \nonumber \\ 
&-  \langle \delta  \hat{\Upsilon} \delta \vec{\bf X} \rangle
   + \frac{1}{l_H^2} \hat{\bf z}  \times \langle \delta  \vec{\gamma}  \delta \bf u \rangle. \label{eq:column_constitutive_H_average}
\end{align}
In keeping with our approximation, the spatial variations, $\delta \mathbf{u}$ and $\delta \vec{\mathbf{X}}$ may be evaluated in terms of $\mathbf{u}_0$  and $\vec{\mathbf{X}}_0$ to zeroth order accuracy in the magnetic field. However the relation between ${\bf u}_0$  and $\vec{\bf X }_0$ is now given by Eq.~(\ref{eq:u_0_X_0_H_result}). Substituting the linear response solutions from Eq.~(\ref{eq:solution_summary_1}) into Eq.~(\ref{eq:column_constitutive_H_average}), we can express the macroscopic particle and entropy current in terms of ${\bf u}_0$ and $\vec{\bf X}_0$ in the form
\begin{align}\label{eq:average_current_u_X_1_H}
 \langle \vec{\bf J}\rangle
  &=\left[ \vec{x}_{\mathrm{e}}  +   \frac{1}{l_H^2} \vec{\gamma}_0 \hat{\bf{z}}  \times   \right] {\bf u}_0  \nonumber \\ 
  &-  \left[ \hat{\Upsilon}_0   + \int \frac{d^2 q}{ 8 \pi^2}  \frac{ 1 }{\eta q^2 } \vec{x}_{-\q} \otimes \vec{x}_\q^{\mathbb{T}} \right] \vec{\bf X}_0. 
\end{align}
Keeping in mind that the average velocity ${\bf u}_0$ is not independent of the thermodynamic force $\vec{\bf X}_0$, but is related by Eq.~(\ref{eq:u_0_X_0_H_result}), we obtain the relation between the macroscopic vector current and the driving forces $\vec{\bf X}_0$,
\begin{align}\label{eq:average_current_u_X_2_H}
 \langle \vec{\bf J}\rangle &
  = -   \frac{\left[(k+ k_H)  +  \frac{n_0 }{l_H^2}  \hat{\bf{z}} \times\right] }{(k+k_H)^2  + n_0^2/l_H^4}  \left[  \vec{x}_{\mathrm{e}} \otimes \vec{x}_{\mathrm{e}}^{\mathbb{T}} - \frac{1}{l_H^4}  \vec{\gamma}_0 \otimes \vec{\gamma}_0^{\mathbb{T}}\right] \vec{\bf X}_0 \nonumber \\
  &  -   \frac{\left[- \frac{n_0 }{l_H^2} + (k+k_H)\hat{\bf{z}} \times\right] }{l^2_H[(k+k_H)^2  + n_0^2/l_H^4]}  \left[  \vec{x}_e \otimes \vec{\gamma}_0^{\mathbb{T}} +  \vec{\gamma}_0 \otimes  \vec{x}_e^{\mathbb{T}}\right] \vec{\bf X}_0
  \nonumber \\
 &
  -  \left[ \hat{\Upsilon}_0   + \int_{\bf q}  \frac{1}{2\eta q^2 } \vec{x}_{-\q} \otimes \vec{x}_\q^{\mathbb{T}} \right] \vec{\bf X}_0.
\end{align}
Equation (\ref{eq:average_current_u_X_2_H}) is the central result that allows us to extract various kinetic coefficients of interest. 


\section{Graphene near charge neutrality}\label{sec:graphene}

We shall apply the main results to study the field dependence of the kinetic coefficients of graphene near the charge neutrality point. The applicability of our long range approximation translates to $\xi \gg l_T\equiv v/T$, where $v$ is the band velocity of graphene and $\xi$ is the correlation length of the disorder potential. One can further make simplifications to $k$ and $\vec{{x}}_e$ by neglecting terms that are small in the ratio $n_0/s_0$. We get
\begin{equation}
	k=\frac{e^2}{\sigma_0}\frac{\langle \delta n^2\rangle}{2},\quad 
	\vec{x}_{\rm{e}} = \left(
                \begin{array}{c}
                  n_0 \\
                  s_0 - \frac{e^2}{2 \sigma_0 T} \langle  \delta n \delta \gamma\rangle \\
                \end{array}
              \right).
\end{equation}
The constraint on the number density, Eq.~(\ref{eq:n_constraint}), reads,
\begin{equation}\label{eq:n0-k-condition}
	|n_0| \lesssim \frac{\xi}{l_T}\sqrt{\langle\delta n^2\rangle}\,.
\end{equation}

We are now prepared to evaluate the magnetotransport coefficients. To this end, by rewriting Eq.~(\ref{eq:average_current_u_X_2_H}) in the form Eq. (\ref{eq:Upsilon_eff_def}), we can easily infer the matrix elements of the conductivity matrices. Indeed, for $(\hat{\Upsilon})_{ij}$ they read 
\begin{subequations}
\begin{align}
&\Upsilon_{11}= \frac{\sigma_0}{e^2}
\left[1+\chi_\sigma+\frac{(x^2-h^2\Lambda)}{\Sigma}\right],\\
&\Upsilon_{12}=\Upsilon_{21}=\frac{n_0s_0}{k\Sigma}, \\
&\Upsilon_{22}=\frac{\kappa}{T}+\frac{s^2_0(1+h^2)}{k\Sigma},
\end{align}
\end{subequations}	
and for $(\hat{\Xi})_{ij}$ they read
\begin{subequations}
\begin{align}
&\Xi_{11}=\frac{\sigma_0}{e^2}\frac{xh}{\Sigma}(1+\Lambda),\\
&\Xi_{12}=\sqrt{\frac{\sigma_0}{e^2}\frac{s^2_0}{k}}\frac{h\Lambda}{\Sigma},\\ &\Xi_{22}=\frac{s^2_0}{k}\frac{xh}{\Sigma}. 
\end{align}
\end{subequations}
Here we introduced several dimensionless parameters to quantify particle density $(x)$, characteristic strength of magnetic field $(h)$, and measure of viscous effects $(\chi)$ normalized in units of density variation induced by long-range disorder, defined respectively as follows:
\begin{align}\label{eq:chi}
x=\frac{\sqrt{2}n_0}{\sqrt{\langle\delta n^2\rangle}},\qquad  h= \frac{\sigma_0}{e^2}\frac{\sqrt{2}}{l_H^2 \sqrt{\langle \delta n^2 \rangle}},
\nonumber \\ 
\chi=\frac{1}{2\eta}\int_{\bf q}\frac{|n_{\bf q}|^2}{q^2},\qquad \chi_\sigma=\frac{e^2\chi}{\sigma_0}.
\end{align}
To make expressions more compact we also introduced two dimensionless functions \begin{subequations}
\begin{align}
&\Sigma(x,h)=(1+h^2)^2+x^2h^2,\\ 
&\Lambda(x,h)=1+x^2+h^2.
\end{align}
\end{subequations}
In order to understand the dependence of these quantities on temperature, it is useful to recall that, modulo logarithmic renormalizations, the intrinsic conductivity $\sigma_0$ is practically a constant \cite{Mishchenko,Kashuba}. Similarly, near charge neutrality, and up to logarithmic factors, shear viscosity is given by $\eta\sim (T/v)^2$ \cite{Fritz}. In order to estimate the friction coefficient $k$, we apply the linear screening approximation that enables one to relate the equilibrium density modulation to the external potential, $n_{\mathbf{q}}=-\nu qU_{\mathbf{q}}/(q+r^{-1})$, where $r=1/(2\pi e^2\nu)$ is the Thomas-Fermi screening radius and $\nu$ is the thermodynamic single-particle density of states. In the hydrodynamic regime, correlation radius of disorder $\xi$ exceeds the Thomas-Fermi screening radius $r$, since $\nu\sim T/v^2$, near charge neutrality, and typical wavenumbers are of order $q\sim \xi^{-1}$. As a result, for the spectral density of disorder potential that does not have strong divergence at $q\to0$, e.g., encapsulation-induced disorder, we therefore get $k\sim (e^2/\sigma_0)\langle U^2\rangle/\xi^2e^4$. In the same regime the entropy density is of the order $s_0\sim(T/v)^2$. It is important to stress that our long wavelength approximation is satisfied for the typical values of $U$ and $\xi$ in the temperature range of hydrodynamic behavior. Indeed, scanning probes on boron nitride encapsulated graphene reveal that the typical correlation radius of density fluctuations (electron-hole charge puddles) is somewhere in the range $\xi\sim 100$ nm and local disorder potential is in the range of $U\lesssim 5$ meV
\cite{Yacoby-Puddles,Crommie,STM-GhBn}. Finally we note that dimensionless parameter $h$ of magnetic field strength can be large, $h\gg1$, already at relatively small fields. 

\begin{figure}[t!]
  \centering
  \includegraphics[width=0.95\linewidth]{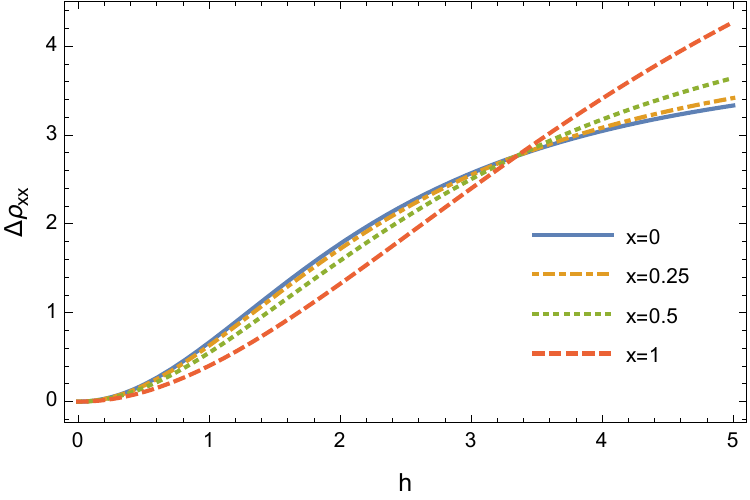}
 \caption{Magnetic field dependence of the relative MR at charge neutrality $x=0$ and at several finite densities $x=0.25, 0.5, 1$ plotted in units of Eq. \eqref{eq:chi}.}
  \label{Fig-drxx-h}
\end{figure}

\subsection{Magnetoresistance and Hall effect}

With these results at hand, we determine the longitudinal $\sigma_{xx}$ and Hall $\sigma_{xy}$ components of the electrical conductivity tensor. In the  notations introduced above, these take the form   
\begin{subequations}\label{eq:sigma-xx-xy}
\begin{align}
	\sigma_{xx}(n_0,H)=e^2\Upsilon_{11}(n_0,H), 
 \\ \sigma_{xy}(n_0,H)=e^2\Xi_{11}(n_0,H).
\end{align}
\end{subequations}
At zero magnetic field, the conductivity reduces to
\begin{equation}\label{eq:sigma-CNP}
	\sigma_{xx}(n_0,H=0)=\sigma_0(1+\chi_\sigma+x^2),
\end{equation}
where the correlational effect of the long range disorder \textit{enhances} the conductivity at charge neutrality by $e^2\chi$ \cite{LLA}.

It is instructive to look at magnetoconductivity precisely at charge neutrality,
\begin{equation}\label{eq:conductivity_charge_neutrality_H}
  \sigma_{xx}(n_0=0,H) =  \sigma_0\left[ \chi_\sigma +\frac{1}{1+h^2}\right].  
\end{equation}
We see that in a  narrow interval of magnetic fields $h \sim 1$  the conductivity decreases from its zero field value $\sigma_0 (1+\chi_\sigma)$ to a plateau at $\sigma_0 \chi_\sigma$. We note that the plateau value $\sigma_0 \chi_\sigma$ coincides with the magnitude of the disorder enhancement of zero-field conductivity at charge neutrality over its intrinsic value. 
Thus magnetoconductivity measurements  may enable determination of the disorder enhancement and provide an additional probe of density variations $\langle\delta n^2\rangle$.   

\begin{figure*}[t!]
  \centering
  \includegraphics[width=0.45\linewidth]{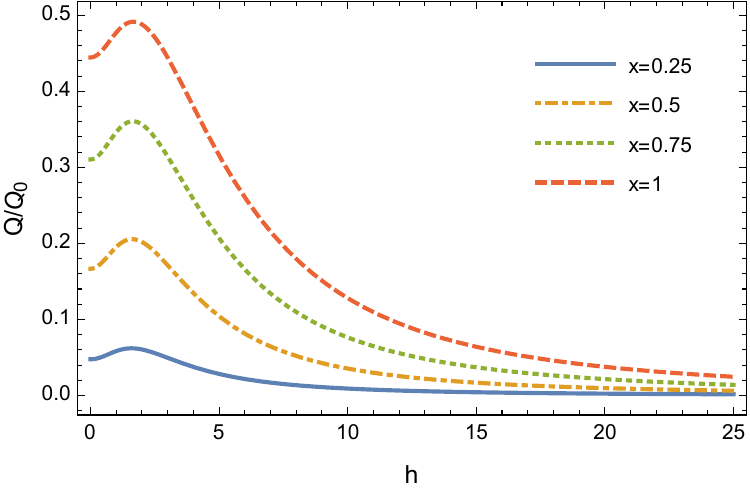}
  \includegraphics[width=0.45\linewidth]{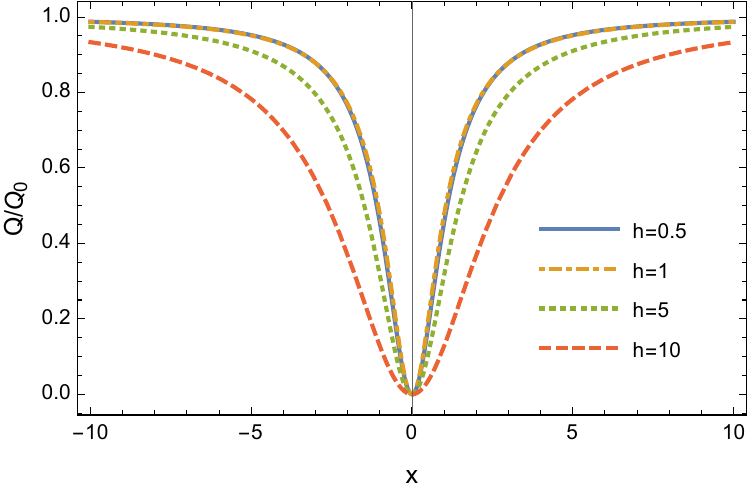}
    \includegraphics[width=0.45\linewidth]{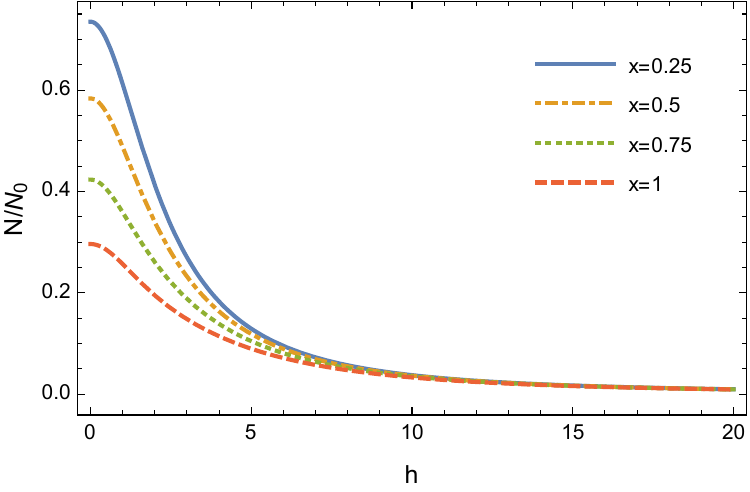}
      \includegraphics[width=0.45\linewidth]{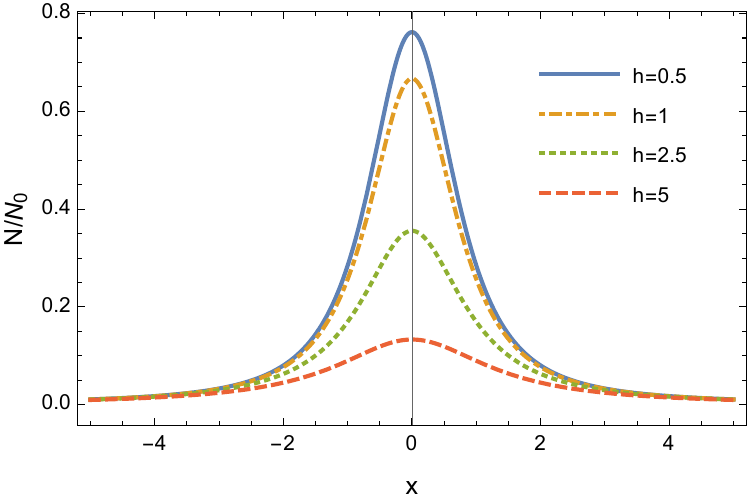}
  \caption{Magnetic field and density dependence for the Seebeck $Q(n_0,H)$ and the Nernst $N(n_0,H)$ transport coefficients. The normalization units are respectively $Q_0=s_0/en_0$ and $N_0=s_0/k$. The choice of parameter $\chi_\sigma$ is the same as in Fig. \ref{Fig-r-xx-xy}.}
  \label{Fig-N-Q}
\end{figure*}

Inverting the conductivity tensor, the longitudinal and Hall resistivity are
\begin{equation}
\rho_{xx}=\frac{\sigma_{xx}}{\sigma^2_{xx}+\sigma^2_{xy}},\quad \rho_{xy}=\frac{\sigma_{xy}}{\sigma^2_{xx}+\sigma^2_{xy}}.
\end{equation}
Substituting the expressions for $\sigma_{xx}$ and $\sigma_{xy}$, and retaining the leading in $H$ contributions we get 
\begin{subequations}
\begin{align}
	& \Delta\rho_{xx}\approx\frac{1+\chi_\sigma+\chi_\sigma x^2(3+x^2)}{(1+\chi_\sigma+x^2)^2}h^2, \label{eq:rho_xx}\\
	& \rho_{xy}\approx\rho_H\frac{x^2(x^2+2)}{(1+\chi_\sigma+x^2)^2}, \label{eq:rho_xy}
\end{align}
\end{subequations}
where $\rho_H=H/(n_0ec)$ is the classical Hall resistance and the relative magnetoresistivity is defined as 
\begin{equation}
\Delta \rho_{xx}(n_0, H)\equiv\frac{\rho_{xx}(n_0, H)-\rho_{xx}(n_0, H=0)}{\rho_{xx}(n_0, H=0)}. 
\end{equation}
 
In the limit of charge neutral graphene, $n_0\to0$, the  above expression simplifies to 
\begin{equation}
\Delta\rho_{xx}(H)\simeq2\left(\frac{\sigma_0}{e^2}\right)^2\frac{\sigma_0}{\sigma_0+e^2\chi}\frac{1}{\langle\delta n^2\rangle l^4_H}. 
\end{equation}
This result is qualitatively consistent with the experimental observations of Ref. \cite{Ponomarenko} as it features strong relative magnetoresistivity. It also shows that the effect is absent in the Galilean invariant system where $\sigma_0\to0$. Furthermore, Eq. \eqref{eq:rho_xx} shows that the effect is suppressed at higher density in agreement with the experiment. Conversely, we can consider the high-density regime $n^2_0\gg \langle \delta n^2 \rangle$, 
which is still consistent with the applicability condition set by Eq. (\ref{eq:n0-k-condition}) as we always require $\xi\gg l_T$. In this case, in accordance with Eq. (\ref{eq:sigma-CNP}), the zero field resistivity $\rho_{xx}(n_0)\approx \sigma^{-1}_0\langle\delta n^2\rangle/2n^2_0$. Then using an estimate for $\chi\sim \xi^2\langle\delta n^2\rangle/\eta$ and an expression for $k$ we arrive at MR in the form 
\begin{align}
\Delta\rho_{xx}(H) 
\simeq\left(\frac{\sigma_0}{e^2}\right)\frac{2\xi^2}{\eta l^4_H}, 
\end{align}
This reproduces an earlier result derived for the case of Galilean invariant electron liquids~\cite{LXA}, where the temperature dependence of MR was predicted to follow inverse proportionality with the fluid viscosity. We add that the sign of MR was argued to remain positive across the ballistic-to-hydrodynamic crossover \cite{Mandal}. It is worthwhile noticing that the Hall coefficient in Eq.~\eqref{eq:rho_xy} is also strongly modified by frictional viscous effects which are the most pronounced near charge neutrality. For completeness in Fig. \ref{Fig-r-xx-xy} we plot both MR and Hall resistivity in a broader range of parameters. In order to highlight the magnitude of the resulting response, we plot separately relative magnetoresistivity $\Delta\rho_{xx}$ at the charge neutrality and at several densities away from neutrality in Fig. \eqref{Fig-drxx-h}. It is clear from the plot that the effect becomes anomalously strong, $\Delta\rho_{xx}\sim1$, already at relatively small fields,  $h\sim1$.  

\subsection{Thermomagnetic phenomena}

We proceed to investigate thermal magnetotransport phenomena. Setting the electric current to zero, we get the relation between the electric field and the temperature gradient from the first row of Eq.~(\ref{eq:Upsilon_eff_def})
\begin{align}\label{eq:EgradientT}
	e \langle \boldsymbol{\mathcal{E}} \rangle = & \, \frac{\Upsilon_{11}\Upsilon_{12}+\Xi_{11}\Xi_{12}}{\Upsilon^2_{11}+\Xi^2_{11}}\,\langle \boldsymbol{\nabla}T \rangle\nonumber \\ 
 &+ \frac{\Upsilon_{11}\Xi_{12}-\Xi_{11}\Upsilon_{12}}{\Upsilon^2_{11}+\Xi^2_{11}} \,\hat{\bf z} \times \langle \boldsymbol{\nabla}T \rangle.
\end{align}
From the first term in the RHS of Eq.~(\ref{eq:EgradientT}), the magnetothermal power or the Seebeck coefficient $Q=\langle\mathcal{E}_x\rangle /\langle\nabla_xT\rangle$ can be expressed as
\begin{align}\label{eq:Seebeck_H}
	&Q(n_0,H) = \frac{1}{e}\frac{\Upsilon_{11}\Upsilon_{12}+\Xi_{11}\Xi_{12}}{\Upsilon^2_{11}+\Xi^2_{11}}\nonumber \\
	&\approx\frac{Q_0x^2}{1+\chi_\sigma+x^2}\left[1+\frac{(1+\chi_\sigma)(1-2\chi_\sigma)-\chi_\sigma^2x^2}{(1+\chi_\sigma+x^2)^2}h^2\right],
\end{align}
where $Q_0=s_0/(e n_0)$ is the value (entropy per unit charge) for the pristine electron liquid and we retained terms only to the leading $H^2$ correction. From the second term in the r.h.s of Eq.~(\ref{eq:EgradientT}), the Nernst coefficient $N=\langle\mathcal{E}_y\rangle /\left(H\langle\nabla_xT\rangle\right)$ is given by
\begin{equation}\label{eq:Nernst_H}
N(n_0,H)= \frac{1}{eH} \frac{\Upsilon_{11}\Xi_{12}-\Xi_{11}\Upsilon_{12}}{\Upsilon^2_{11}+\Xi^2_{11}}.
\end{equation}
At charge neutrality, $n_0\to0$, matrix elements of $\hat{\Upsilon}_{ij}$ and $\hat{\Xi}_{ij}$ simplify greatly and we obtain
\begin{equation}
N(H)=\frac{\sigma_0}{e^2}\frac{2s_0}{\langle\delta n^2\rangle}\frac{1}{1+\chi_\sigma(1+h^2)}.
\end{equation}
The most significant implication of this result is that in systems with broken Galilean invariance, the Nernst coefficient remains finite at charge neutrality, which is made possible by the finite intrinsic conductivity. However, as is clear from the above expression, disorder is crucial to stabilize the result. The field dependence is that of a Lorentzian shape. For clarity, we plot $Q$ and $N$ on Fig. \eqref{Fig-N-Q} as a function of magnetic field for several different values of electron density and then as a function of density for different strengths of the field.      

\begin{figure*}
  \centering
  \includegraphics[width=0.45\linewidth]{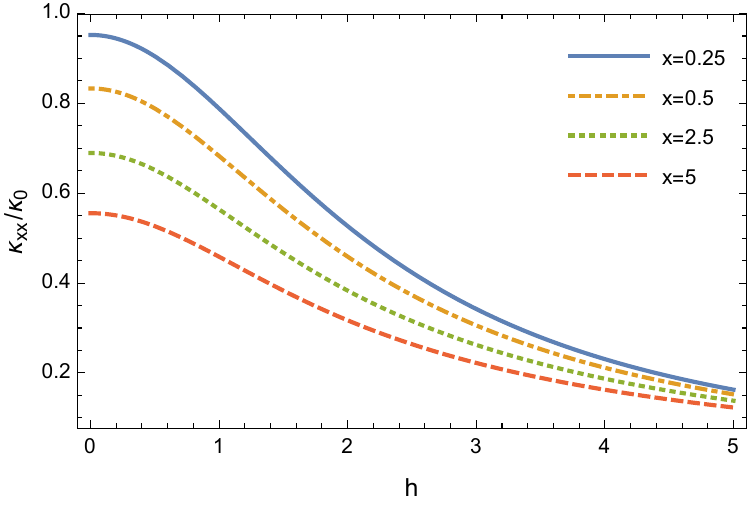}
  \includegraphics[width=0.45\linewidth]{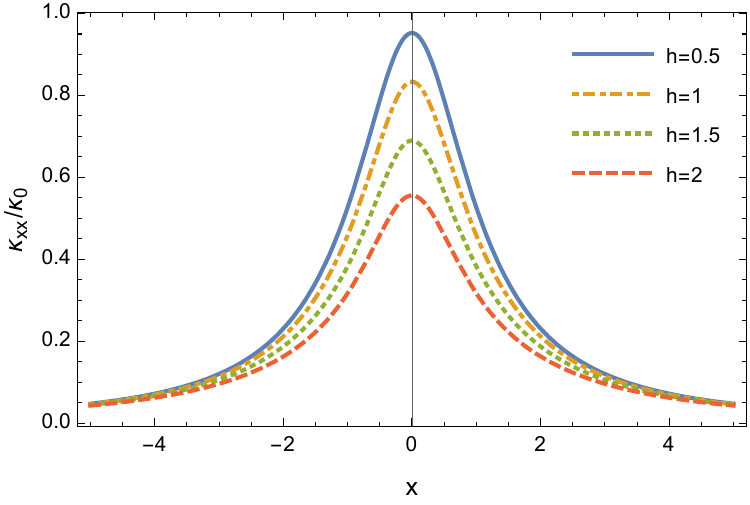}
   \includegraphics[width=0.45\linewidth]{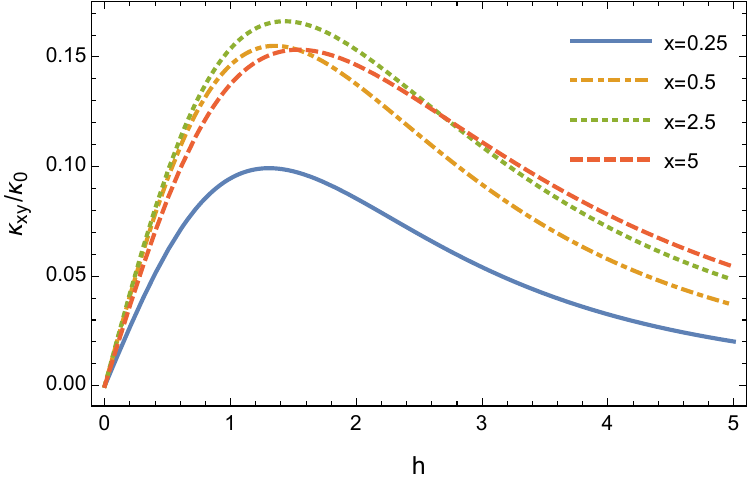}
  \includegraphics[width=0.45\linewidth]{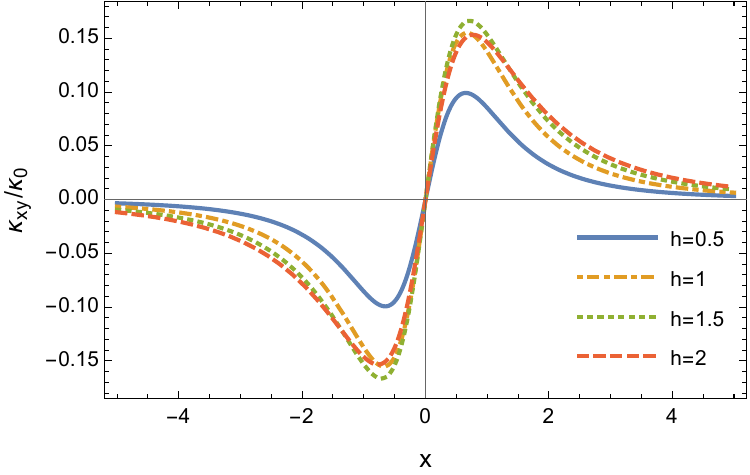}
   \caption{Magnetic field and density dependence for the longitudinal $\kappa_{xx}(n_0,H)$ and Hall $\kappa_{xy}(n_0,H)$ thermal conductivities normalized in units of $\kappa_0=Ts^2_0/k$. The choice of parameter $\chi_\sigma$ is the same as in Fig. \ref{Fig-r-xx-xy}.}
  \label{Fig-k-xx-xy}
\end{figure*}

In a similar fashion, substituting the expression for the electric field, namely Eq.~\eqref{eq:EgradientT}, into the second row of Eq.~\eqref{eq:Upsilon_eff_def}, one finds the thermal conductivity and thermal Hall conductivity in terms of the matrix elements of $\hat{\Upsilon}$ and $\hat{\Xi}$,
\begin{subequations}
\begin{align}
	&\frac{\kappa_{xx}(n_0,H)}{T} =  \Upsilon_{22} - \frac{\Upsilon_{11}\left(\Upsilon^2_{12}-\Xi^2_{12}\right)+2\Upsilon_{12}\Xi_{11}\Xi_{12}}{\Upsilon^2_{11}+\Xi^2_{11}}, \\
	&\frac{\kappa_{xy}(n_0,H)}{T} = -\Xi_{22} + \frac{\Xi_{11}\left(\Xi^2_{12}-\Upsilon^2_{12}\right)+2\Xi_{12}\Upsilon_{11}\Upsilon_{12}}{\Upsilon^2_{11}+\Xi^2_{11}}.
	\end{align}
\end{subequations}
In the following, we focus on the limiting cases of vanishing field at finite density and finite field at charge neutrality, where analytical expressions greatly simplify 
\begin{subequations}
\begin{eqnarray}
\kappa_{xx}(n_0,H=0) &=&\kappa_0 \frac{1+\chi_\sigma}{1+\chi_\sigma+x^2},\\ \kappa_{xx}(n_0=0,H)&=&\kappa_0\frac{1+\chi_\sigma}{1+\chi_\sigma(1+h^2)}. 
\end{eqnarray}
\end{subequations}
Here $\kappa_0=Ts^2_0/k$: see Refs. \cite{LLA,Lucas-PRB16}. In these formulas we neglected the term in $\Upsilon_{22}$ with intrinsic coefficient since $\kappa\ll Ts^2_0/k$, so the thermal transport is dominated by disorder contribution. In both cases thermal conductivity has a Lorentzian shape either as a function of $n_0$ or $H$. It is evident that the correction term to the diagonal thermal conductivity at small fields is negative. It can be readily shown that this remains true at all densities. In particular, at high density limit, $n^2_0\gg k$, the leading correction in $H^2$ takes the form 
\begin{equation}
	\Delta \kappa_{xx} \equiv \frac{\kappa_{xx} (n_0,H)-\kappa_{xx} (n_0,0)}{\kappa_{xx} (n_0,0)}\approx - (1-\chi_\sigma+\chi_\sigma^2)
	\frac{h^2}{x^2}.
\end{equation}
For the thermal Hall conductivity, also known as the Righi–Leduc effect, we get at leading order (linear in $H$)
\begin{equation}
	\kappa_{xy}(n_0,H) = \kappa_0 \frac{xh(1-\chi_\sigma^2)}{(1+\chi_\sigma+x^2)^2}. \label{eq:thermal_Hall}
\end{equation}
At weak field the subsequent corrections can be organized in powers of $1/kl^4_H$ and it can be readily checked that the next term is negative. Figure \eqref{Fig-k-xx-xy} summarizes density and field dependence of the longitudinal and Hall thermal magnetoconductivity.  


\section{Discussion of the results}

\label{sec:summary}

In this paper we have constructed a hydrodynamic description of magnetotransport for conductors lacking Galilean invariance in the presence of long-ranged inhomogeneity. While the effect of disorder on transport is in general nonperturbative in this regime, we provided analytical solutions to various transport coefficients exploring the limit of smooth disorder with the focus on graphene near charge neutrality. Our consideration shows that, near charge neutrality, the MR is positive and quadratic in field. However, the quadratic coefficient acquires a different dependence on density, shear viscosity, and correlation length of disorder as compared to that in Ref.~\cite{LXA}. Furthermore, the saturation of MR at higher fields enables separation of intrinsic conductivity and viscous contribution mediated by disorder. The Hall component of the resistivity is given by the product of the classical Hall resistance and a renormalization factor, which is a function of doping density, shear viscosity, and correlation length of disorder. The effect of viscosity and disorder on thermal-magnetohydrodynamic transport is also analyzed. In particular, thermal conductivity is strongly enhanced by disorder on top of its intrinsic value. This effect diminishes in the magnetic field. The Nernst coefficient and thermal power are strongly affected by viscous effects and long-range disorder potential. Interestingly, the Seebeck coefficient exhibits nonmonotonic dependence on the field with a pronounced maximum that can be probed experimentally. We hope our results can stimulate further experimental investigations in magnetohydrodynamic transport in electronic systems, especially graphene based devices near charge neutrality.

We note that the dependence of the transport coefficients on the magnetic field is described by the dimensionless parameter $h$, which becomes of order unity at relatively weak magnetic fields, whose scale is set by the intrinsic conductivity and  disorder strength, $1/l_H^2 \sim \frac{e^2}{\sigma_0}\sqrt{\langle \delta n^2 \rangle}$. This sensitive dependence on the magnetic field arises from the competition between the disorder-induced and magnetic friction forces discussed below Eq.~\eqref{eq:x_e_long_wavelength}. At $h \lesssim 1$ the friction force is dominated by disorder, while for $h \gtrsim 1$ magnetic friction becomes dominant. For a quantitative estimate we can assume that intrinsic conductivity $\sigma_0/e^2\sim1$ and take density variation in the range $\delta n\sim 10^9 - 10^{10}$ cm$^{-2}$. This gives $h\sim 1$ for the field strength of the order of $\sim 10 -100$ mT, respectively. Already at such small fields resistance increases by 100\% leading to the giant effect. 

The mechanism of the anomalously strong magnetotransport can be qualitatively understood by considering magnetoconductivity at charge neutrality, Eq.~\eqref{eq:conductivity_charge_neutrality_H}. In this case the applied electric field does not produce the average force. Therefore the average flow velocity is determined by the competition between the two friction forces. At small magnetic fields $h\ll 1$ the drift velocity is negligible in comparison to $c[\mathbf{E}\times {\bf H}]/H^2$. Therefore, the EMF (which is evaluated in the liquid frame) is nearly equal to $e\mathbf{E}$ and we get the zero field result. At $h\gg 1$ disorder-induced friction is relatively small and the fluid velocity approaches $c[\mathbf{E}\times {\bf H}]/H^2$. In this case the EMF vanishes and the contribution of the intrinsic conductivity [second term in Eq.~\eqref{eq:conductivity_charge_neutrality_H}] is suppressed. On the other hand, the nonuniform force on the liquid, which is proportional to local electron density and the electric field is not affected by the magnetic field. Therefore, the disorder enhancement of the zero field conductivity, which is described by the first term in Eq.~\eqref{eq:conductivity_charge_neutrality_H}, remains unchanged. In closing we note that our consideration focused on the bulk contribution to magnetotransport coefficients where momentum relaxation is driven by the disorder potential and the flow velocity is uniform.  In finite samples where momentum relaxation occurs both in the bulk and at the sample boundaries a similar physical mechanism will also play a role in magnetotransport. An extension of the present theory to finite samples with Hall bar and Corbino geometries will be presented elsewhere~\cite{Zverevich}.    


\section*{Acknowledgments}

We thank A. L. Berdyugin, I. Gornyi, P. Kim, L. Levitov, A. Lucas, B. Narozhny, L. A. Ponomarenko, B. Skinner, A. Talanov, and J. Waissman
for useful discussions on various topics related to this work. This work (A.~L. and S.~L.) was supported by the National Science Foundation Grant No. DMR-2203411 and H. I. Romnes Faculty Fellowship provided by the University of Wisconsin-Madison Office of the Vice Chancellor for Research and Graduate Education with funding from the Wisconsin Alumni Research Foundation. A.~V.~A. acknowledges support from the National Science Foundation through the MRSEC Grant No. DMR-1719797, the Thouless Institute for Quantum Matter (TIQM), and the College of Arts and Sciences at the University of Washington. A.~L. acknowledges the hospitality of the Kavli Institute for Theoretical Physics (KITP), where this paper was written in part, which is supported by the National Science Foundation under Grants No. NSF PHY-1748958 and No. PHY-2309135.



\begin{thebibliography}{100}
\bibitem{Note}
The universality of the coefficient $\frac{e}{c}$ in Eq. \eqref{eq:X'_def} is a consequence of the gauge invariance. Similar considerations lead Ginzburg and Landau to conclude that the effective charge $e^*$ in the Ginzburg-Landau equations cannot be a material-dependent parameter \cite{Ginzburg-Landau}. Otherwise an inhomogenous deformation of the crystal would violate gauge invariance.
\end{thebibliography}
\end{document}